\documentstyle[times,pramana,epsf,graphicx,rotating,floats]{ias}

\begin{document}
\mark{{Novel interference}{P. Singha Deo and A. M. Jayannavar}}
\title{Novel interference effects and a new Quantum phase in
  mesoscopic systems }

\author{P. Singha Deo\cite{psd}}
\address{S. N. Bose National Center for Basic Sciences, JD Block,
Sector III, Salt Lake City, Calcutta, India}

\author{ A. M. Jayannavar\cite{amj}}
\address{Institute of Physics, Sachivalaya Marg, Bhubaneswar - 751
  005, India}


\pacs{2.0}

\abstract{

Mesoscopic systems have provided an opportunity to study quantum effects
beyond the atomic realm. In these systems quantum coherence prevails over
the entire sample. We discuss several novel effects related to persistent
currents in open systems which do not have analogues in closed systems.
Some phenomena arising simultaneously due to two non-classical effects
namely, Aharonov-Bohm effect and quantum tunneling are presented. Simple
analysis of sharp phase jumps observed in double-slit Aharonov-Bohm
experiments is given. Some consequences of parity violation are
elaborated. Finally, we briefly describe the dephasing of Aharonov-Bohm
oscillations in Aharonov-Bohm 
ring geometry due to spin-flip scattering in one of the
arms. Several experimental manifestations of these phenomena and their
applications are given.

}

\maketitle

\section{Introduction}

Mesoscopic physics deals with samples that are intermediate in size
between the atomic scale and the macroscopic scale determined by the
transport phase coherence length of electrons or quasiparticles. Studies
in these systems have revealed a new range of unexpected quantum
phenomena, often counter-intuitive. Interpretation of these phenomena
requires full recognition of the wave nature of quasiparticles and keeping
track of their phase coherence over the entire sample including the
measurement leads and probes (quantum measurement process). Clearly, when
the transport dimension reaches the charge-carrier inelastic scattering
length (coherence length) and charge confinement dimension approaches
Fermi wavelength, then the physics of these systems, based on the motion
of particles and ensemble averaging is expected to be invalid. The notion
of the usual ensemble averaged transport coefficient such as the
resistivity/conductivity, that is local and material specific, has to be
replaced by that of resistance/conductance, that is global and
operationally specific to the sample as well as the nature of probe for
measurements.

The guiding theme of mesoscopic physics is the quantum interference over
the entire sample or treating the whole mesoscopic sample as a single
quantum scatterer. These systems exhibit properties where interference of
electronic waves, quantization of energy levels (quantum size effects),
discreteness of charge and their number being even or odd (parity effects)
play a major role. Thus we have an opportunity for exploring truly quantum
effects beyond the atomic realm. This subject enjoys the unique position
of being able to deal with and provide answers on fundamental questions in
physics while being relevant to applications in the area of quantum
electronics, computation and communication which are rapidly emerging
fields in their own respect. Basic questions about how the quantum rules
operate and go over to classical ones at the macroscopic level as one
tunes the temperature or sample size are being answered, i.e., mesoscopic
systems present the possibility of studying in a controlled way the
process of decoherence (or dephasing) and the transition from quantum to
classical behavior.

Some of the experimentally observed phenomena in these systems include
breakdown of classical Ohm's law \cite{wha}, the normal state
Aharonov-Bohm (AB) oscillations in resistance \cite{web}, normal electron
persistent currents \cite{lev}, non-local current and voltage relations
(manifestation of quantum non-locality) \cite{was}, conductance
quantization of point contacts \cite{wha}, normal and anomalous quantum
Hall effect, negative four probe resistances, quantum shot noise, single
electron transistor (Coulomb blockade), proximity effect in mesoscopic
superconductors, spin coherence effects (spintronics), entangled states in
quantum dots etc. to name but a few. Interestingly, many of these
phenomena can be observed through the use of straight forward experimental
probes, namely the dc electrical two probe and four probe conductance in
the presence or absence of magnetic field.

The propagation of electron has many interesting similarities with other
wave propagation such as electromagnetic wave or sound wave propagation
etc. and thus mesoscopic phenomenon can be understood in terms of wave
guide theory. In this work we will discuss various mesoscopic phenomena
that we studied using the mode matching technique for wave guides. Such a
study can also be performed using Greens function techniques or using the
tight binding Hamiltonian, but for non-interacting systems the mode
matching technique allows more explicit and exact calculations. Quantities
like conductance, local and global currents and eigenenergies can be
calculated from first principles. Most of the phenomena discussed here are
new and were pointed out by us. Some of them have experimental
consequences as well as applications in quantum devices.

In the following sections we discuss several novel effects related to
persistent currents and transport in open systems (connected to external
reservoirs), which have no analogue in closed or isolated systems. Some of
the effects which we discuss arise simultaneously due to two non-classical
effects namely AB-effect and quantum tunneling. We also discuss the
observed additional new phase of electron wavefunction in the AB-ring
geometry, breakdown of parity effects and their consequences. Finally we
briefly discuss our ongoing work on dephasing of AB-oscillations in the
presence of a magnetic impurity (leading to exchange spin flip scattering)
in one of the arms of the AB-ring. This has resemblance with which-way
interferometer or which-path detector models developed to study dephasing
due to quantum measurement process.

\section{Directional dependence of persistent currents and the quantum
current magnification effect}

Persistent currents in closed loops are equilibrium quantum mechanical
currents that were first predicted \cite{but1} and subsequently detected
in normal metals \cite{lev} and semiconductor \cite{mai} rings.  Such
currents were theoretically known for a very long time \cite{bye}.  The
application of magnetic field (AB flux) destroys the time reversal
symmetry and as a consequence persistent currents flow in the loop and are
periodic in magnetic flux, with a period $\phi_0$, $\phi_0=hc/e$ being the
elementary flux quanta. For a perfect ring of circumference $L$ the
magnetic flux $\phi$ enclosed by the loop modifies the periodic boundary
condition into $\psi(x+L)=\psi(x)e^{i2\pi\phi/\phi_0}$. This tuning of
boundary condition changes the energy levels in a periodic manner.
Persistent current, being the flux derivative of the total free energy, is
also periodic in flux. Thus persistent current can be attributed to
changing boundary conditions due to magnetic flux. At zero temperature for
spinless electrons persistent current is diamagnetic or paramagnetic
depending on the number of electrons in the loop being odd or even
respectively (parity effect).  It was subsequently shown \cite{but2,mai}
that persistent currents can also occur in open systems i.e., when the
system can exchange electrons with electron reservoirs.  In such open
systems, apart from persistent currents one can also have transport
currents. While transport currents are non-equilibrium currents and
require a potential difference (voltage difference, temperature gradients
or chemical potential difference), persistent currents are local internal
currents that do not explicitly depend on the potential difference. We
show that one can distinguish between these two currents using the two
geometries shown in Fig.~1 (a) and (b). These two currents, although exist
simultaneously, are fundamentally different. These geometries help to
probe their different properties.

The quantum mechanical scattering wave functions can be explicitly written
down (depending on the direction of current) in different regions and can
be matched at the junctions using Griffith's boundary conditions
\cite{gri}.  We will be considering for simplicity one dimensional free
electron networks. Appropriate wave functions in the different regions are
thus linear combinations of simple plane wave solutions. The effect of
magnetic field can be incorporated in the boundary conditions. From the
wave functions one can also calculate the currents in different regions.
In case of Fig.~1 (a) one can only have persistent currents in the loop
while the transport currents flow in the wires between reservoirs 1 and 2
at chemical potentials $\mu_1$ and $\mu_2$. In case of Fig.~1 (b) there
will be persistent currents as well as transport currents in the loop. The
two can be separated because persistent current is an odd function of the
flux $\phi$ while the transport current (being proportional to
transmission coefficient)  is an even function of the flux $\phi$. Apart
from being an even function of flux, the transport current also has a flux
independent part but the persistent current by definition has no flux
independent part. Once the two currents are separated using these basic
properties, we find that the magnitude of the transport current is
independent of the direction of the current flow between the reservoirs
(conductance of the sample is the same whether $\mu_1>\mu_2$ or vice
versa). But the magnitude of the persistent current in the loop depends on
the direction of the direct inter-reservoir current.  The defect at the
position X plays a very important role in realizing this directional
dependence of persistent currents. The defect breaks the spatial symmetry
in the problem. Depending on the direction of current flow (for different
scattering problem) we will have a different complex amplitude of wave
function at the junction between loop and the wire.  This amounts to
changing the boundary condition. As discussed above the persistent
currents being sensitive to boundary conditions will have different
magnitudes. For another extreme example when the defect strength becomes
infinite then reservoir 2 is cut off and does not contribute to the
persistent current in the loop. Hence the condition whether $\mu_1>\mu_2$
or not will definitely make a difference. The directional independence of
conductance is a fundamental property of two probe Landauer conductance
and most quantities like noise and fluctuations are always independent of
direction of the current flow. In this respect the properties of
persistent currents in open systems is very unique.

Next we consider a situation when the impurity at site X in Fig.~1 (b)  
and the Aharonov-Bohm flux are not essential but the lengths of the two
arms of the loop $L_1$ and $L_2$ are different. Asymmetry in the two arms
of the ring is a must to obtain this quantum effect. When the two arms are
identical in all respects then half the transport current flows through
the upper arm and the rest half through the lower arm i.e.,
$I/2=I_1=I_2>0$. Here $I$ is the magnitude of the total inter-reservoir
current or the transport current and $I_1$ and $I_2$ are the magnitudes of
currents in the two arms of the ring.  In the case of classical wires when
the arms are not identical then we have the condition $I_1/I_2=R_2/R_1$
and $I=I_1+I_2$ (current conservation or Kirchoff's law) which
automatically implies $0<I_1<I$ and $0<I_2<I$. Here $R_1$ and $R_2$ are
the resistances of the two arms of the ring. Both $I_1$ and $I_2$ are
positive and flow in the same direction of applied fields. But for quantum
wires, a simple scattering solution of the Schrodinger equation shows that
the above conditions break down. At certain Fermi energies and values of
loop parameter one can realize such situations as $I_1>I$ which
automatically requires $I_2<0$, because the current conservation warrants
$I=I_1+I_2$. Thus the current $I_2$ flows against the potential drop. Such
a phenomenon occurs in the vicinity of resonances in the ring and is
quantum mechanical in origin. In classical parallel resonant LCR circuits
(capacitance C connected in parallel with a combination of inductance L
and resistance R) driven by external electromotive force, circulating
currents arise at resonant frequency.  This effect is sometimes referred
to as current magnification. Hence we call the phenomenon discussed above
as ``quantum current magnification effect'' which gives a circulating
current that we define as follows. If both $I_1$ and $I_2$ are positive or
flow in the same direction of the potential drop then the circulating
current is zero. If one of them, say $I_1$ is negative then circulating
current is $I_1$ in both the arms. Of course in the lower arm it is in the
direction of potential drop but not in the upper arm. $I-I_1$ is the
transport current flowing in the lower arm in addition to the circulating
current $I_1$. The parameter regimes where this circulating current exists
can be found in Ref. \cite{jay1}. Experimentally, it is possible to
observe this current magnification effect as a large magnetic response of
the ring by properly tuning either the Fermi energy or other material
parameters. This is due to the fact that magnetic moment of a loop is
proportional to the total integration of the current over the entire loop.
It should be noted that we are obtaining magnetic response in the absence
of applied external magnetic field, however, in the presence of a
transport current. This is a non-equilibrium phenomenon.

Now it can also be shown that unlike the conventional persistent currents,
the quantum current magnification effect can be enhanced by impurity
scattering in certain range of parameter values. To understand that such
an effect is possible consider, for example, a case when the two arms of
the ring are identical in all respect. In that case by symmetry we have
$I_1=I_2=I/2>0$ and hence there is no quantum current magnification
effect. Now if we introduce an impurity potential (say a delta function
potential) in the upper arm of the ring then the symmetry is destroyed and
one can thus in principle obtain a current magnification effect. This
simple example shows that one can have enhancement of current
magnification effect due to the impurity scattering. Although this is not
a general feature but an extremely parameter dependent one, nevertheless
for any configuration of the two arms of the ring one can always find
parameter regimes where the circulating current can be enhanced by
impurity scattering. In the parameter regimes other than these the
circulating current will decrease due to scattering. A detailed analysis
of this can be found in Ref. \cite{par}. We have also shown that there is
no upper bound for the current magnification. This effect has been
extended to thermal currents \cite{moska} and to the spin currents in the
presence of Aharonov-Casher flux \cite{ryu1}.

\section{Persistent currents and transport due to evanescent modes
in the presence of AB-flux}

Let us imagine a geometry where a metallic loop is coupled to a single
electron reservoir via an ideal lead (Fig.~2). In the ideal lead the
potential is assumed to be zero, i.e., V=0. In the metallic loop the
potential throughout the circumference is V and is positive. When injected
electrons have their energies less than V, these electrons can tunnel into
the loop quantum mechanically and propagate inside the loop as evanescent
modes and give rise to a persistent current in the presence of a magnetic
field. Such currents arise simultaneously due to two non-classical
effects, namely, quantum tunneling and Aharonov-Bohm effect. Currents due
to such evanescent modes are to be found by analytical continuation and we
have obtained an analytical expression for these persistent currents
\cite{deo1}.  In the limit QL$>>$1, persistent currents in a small energy
interval around E are given by $dj=f(k,Q)e^{-QL}\sin(\phi/\phi_{0})$,
where f(k,Q) is a simple function of k and Q. Here k is the wave vector
for incident electrons, i.e., k=$\sqrt{2mE/\hbar^{2}}$,
Q=$\sqrt{2m(V-E)/\hbar^{2}}$ and $L$ is the circumference of the loop. As
expected the persistent currents are periodic in magnetic flux with the
period $\phi_{0}$. Owing to the decaying nature of evanescent modes, the
factor arising due to the sensitivity of the wavefunctions to the boundary
conditions appears as $e^{-QL}$. Higher harmonics in magnetic flux (say
nth harmonic) also contribute to the persistent currents with a
multiplication factor $\sin(n\phi/\phi_{0})$. However, these harmonics are
weighted by $e^{-nQL}$ because for these harmonics to appear the electron
has to traverse the loop n times. So, these harmonics can be neglected in
the limit QL$>>$1. Unlike the behavior of persistent currents above the
barrier regime the currents due to evanescent modes do not oscillate as a
function of the Fermi energy as long as E$<$V. The total persistent
current is given by sum of contributions from the electrons up to Fermi
energy. Even though the current due to individual evanescent modes is
small the total sum can have an observable amplitude. Especially in a real
physical situation one can have a ring with extremely narrow width
connected to the reservoir via an ideal wire with a much larger width. In
this situation the zero point quantum potential due to the transverse
confinement in the ring is much higher than the zero point energy of the
ideal wire. Electrons can occupy several subbands in the connecting wire
but still they have energies less than the zero point energy of the ring.
All these electrons in several subband modes will propagate as evanescent
modes in the ring, and in this situation a higher contribution to the
total persistent current may arise.

Having shown that persistent currents due to evanescent modes in open
systems is always diamagnetic, it would be interesting to see if the same
is true in closed systems. One can excite evanescent modes throughout the
circumference of the ring in the following closed system. Consider the
case of one dimensional loop with higher potential $V$ connected to a stub
of length $L$ for which the reference potential is taken to be zero (see
Fig.~3).  For a sufficiently large value of $V$ and long stub there can be
many modes that are propagating in the stub but are classically forbidden
in the ring. However, quantum mechanical tunneling leads to evanescent
modes in the ring. It has been shown that these discrete evanescent modes
always carry a diamagnetic persistent current \cite{deo2}.

As mentioned in the introduction, similarity with the guided
electromagnetic wave propagation has opened up the possibility of new
quantum devices. These devices rely on quantum effects for their operation
and are based on interferometric principles. Several switching devices
have been proposed wherein one can control the relative phase difference
between the two interfering paths by applying electrostatic potentials or
magnetic fields. The transmission across these devices can be varied
between zero and one (100 percent modulation), if the propagation takes
place in the fundamental transverse mode (single channel regime). This
requires that the Fermi energy should be between the ground and the first
excited mode. Otherwise the mode mixing tends to average out the
transmission oscillations. However, the proposed quantum devices are not
very robust in the sense that the operational characteristics depend very
sensitively on the material parameters. Incorporation of a single
impurity, however weak, in the mesoscopic device may change non-trivially
the interference of partial electron waves throughout the sample and hence
electron transmission (operational characteristics across the sample).
Such devices can be exploited only if we achieve the technology that can
reduce or control the phase fluctuations to a small fraction of $2\pi$.

We have studied the transmission across normal metallic loop connected to
reservoirs by ideal wires in the presence of magnetic or AB flux. When
electrons travel as evanescent modes in the loop we find that initial
differential magnetoconductance is always negative and is unaffected by
the presence of impurities in the loop \cite{jay2}. Here transport arises
in the presence of two non-classical effects namely, AB-effect and quantum
tunneling. The above situation can arise in a system in which the
transverse width of the loop is much less than the width of the ideal
wires. Then due to the higher zero point energy arising from transverse
confinement, fundamental subband minima in the loop will be at a higher
energy than the value of the few subband minima in the ideal connecting
wires. Then a situation can arise, where several propagating modes in the
wire will have energy less than the minimum propagating subband energy in
the loop. Thus the electron propagating in a fundamental subband of an
ideal wire feels a barrier to its motion (arising solely due to the
mismatch of the zero point energies) and thus the electron tunnels through
across the loop (due to evanescent propagation) experiencing a higher
effective potential barrier $V$. The fact of initial differential
magnetoconductance being negative can be used for the operation of a
quantum switch where the on and off states would correspond to
transmission in the absence or presence of magnetic field respectively,
i.e., the on state always has larger conductance than the off state. This
difference in conductance can be made large by purposefully incorporating
weak impurities in the arms of the connecting leads so long as they do not
create resonant states in the system \cite{bik}.  The robustness in the
behavior of the initial differential magnetoconductance is related to the
dominance of the first harmonic component in flux in the presence of
evanescent modes only (as in the case of persistent currents discussed
above).  This is in contrast to the fact that the initial differential
magnetoconductance for propagating modes can be either positive or
negative and is very sensitive to small changes in geometric details and
Fermi energy.

\section{A new phase of the electron wave function}

In this section we discuss parity effects and observed quantum phase
slips, a subject of ongoing current interest. States in a ring pierced by
a magnetic flux exhibit strong parity effect \cite{che}. There are two
ways of defining this parity effect in the single channel ring
(multichannel rings can be generalized using the same concepts). In the
single particle picture (possible only in absence of electron-electron
interaction) it can be defined as: states with an even number of nodes in
the wave function carry diamagnetic currents (positive slope of the eigen
energy versus flux) while states with an odd number of nodes in the wave
function carry paramagnetic currents (negative slope of the eigen energy
versus flux) \cite{che}. In the many body picture (without any
electron-electron interaction), it can be defined as: if $N$ be the number
of electrons (spinless) in the ring, the persistent current carried by the
$N$ body state is diamagnetic if $N$ is odd and paramagnetic if $N$ is
even \cite{che}. Leggett conjectured \cite{leg} that this parity effect
remains unchanged in the presence of electron-electron interaction and
impurity scattering of any form. His arguments can be simplified to say
that when electrons move in the ring, they pick up three different kind of
phases: 1) Aharanov-Bohm phase due to the flux through the ring, 2)
statistical phase due to electrons being Fermions and 3) phase due to wave
like motion of electrons depending on their wave vector. The parity effect
is due to competition between these three phases along with the constraint
that the many body wave function satisfy periodic boundary condition
(which means if one electron is taken around the ring with the other
electrons fixed, the many body wave function should pick up a phase of
2$\pi$ in all). Electron-electron interaction or simple potential
scattering cannot introduce any additional phase although it can change
the kinetic energy or the wave vector and hence modify the third phase.
Simple variational calculations showed that the parity effect still holds
\cite{leg}. Multichannel rings can be understood by treating impurities as
perturbations to decoupled multiple channels, which means small impurities
just open up small gaps at level crossings within the Brillouin zone and
keep all qualitative features of the parity effect unchanged.

In a one dimensional (1D) system where we have a stub of length $v$
strongly coupled to a ring of length $u$ (the potential everywhere is
uniform and all modes are propagating modes), we can have bunching of
levels with the same sign of persistent currents \cite{deo2}. This is
essentially because if the Fermi energy is above the value where we have a
node at the foot of the stub (that results in a transmission zero in
transport across the stub), there is an additional phase of $\pi$ arising
due to a slip in the Bloch phase \cite{deo3,sre} (Bloch phase is the third
kind of phase discussed above, but the extra phase $\pi$ due to slips in
Bloch phase is completely different from any of the three phases discussed
above because this phase change of the wave function is not associated
with a change in the group velocity or kinetic energy or the wave vector
of the electron \cite{deo3,sre}). This is illustrated in Fig. 4 (a) and
(b). The y-axis gives the phase of the electron wavefunction inside the
ring, which is nothing but Bloch phase $\cos^{-1}Re[1/t]$. The x-axis
gives the energy $E=k^2$. Fig 4 (b) exhibits discontinuous changes in the
Bloch phase precisely as the energy crosses the transmission zeroes of the
stub. In Fig 4 (a) the length of the stub and the ring is so tuned that
the discontinuous phase slips cannot occur. Fig 4 (a) is similar to Fig 6
in Ref \cite{che} where there is no violation of parity effect. But Fig 4
(b) exhibits violation of parity effect. A more detailed analysis can be
found in Ref. \cite{deo3}. It is worth mentioning that parameter values
for which one can obtain no phase discontinuities as in Fig. 4 (a) forms a
set of measure zero. A physical understanding of the phase slip can be
obtained by mapping the stub to an effective potential of $V(k,x)=k
\cot(kL) \delta (x)$. As the effective potential has discontinuities at
$kL=\pi/2, 3\pi/2,...$, at these energies the scattering phase and hence
the Bloch phase of the electron in the ring will also exhibit
discontinuities.

In an energy scale $\Delta_u\propto 1/u$ (typical level spacing for the
isolated ring of length $u$) if there are $n_b\sim\Delta_u/\Delta_v$
(where $\Delta_v\propto 1/v$, the typical level spacing of the isolated
stub of length $v$)  such phase slips then each phase slip gives rise to
an additional state with the same slope and there are $n_b$ states of the
same slope or same parity bunching together with a phase slip of $\pi$
between each two of them \cite{deo3}. Transmission zeroes is an inherent
property of Fano resonance generically occurring in mesoscopic systems and
this phase slip is believed to be observed \cite{deo4,deo5,ryu2,lee,tan}
in a transport measurement \cite{sch}. The experiment \cite{sch} was done
by embedding a quantum dot in one arm of an Aharonov-Bohm ring. The set up
is an analog of the Young double slit experiment with a mica sheet on the
path of an interfering beam, where one can estimate the phase acquired when
light travels through the mica sheet by observing the change of
interference pattern on the screen. In this case one can estimate the
phase $\theta_d$ acquired by an electron in passing through the dot. The
experiment was first interpreted in terms of Friedel sum rule in Ref.
\cite{ye}. The dot contains many electrons that are strongly interacting
with each other. As the Fermi energy sweeps through a resonance one
additional electron is added to the dot and $\theta_d$ changes by $\pi$ as
is expected from Friedel sum rule. The exact nature of this phase change
could be explained from the theory of Bright-Wigner resonance. Now Friedel
sum rule also suggests that no more $\pi$ phase shift should occur unless
the next electron is added at the next resonance. But between two
resonances an additional change of $\pi$ over a surprisingly small energy
scale was observed in the experiment. This small energy scale is at least
an order of magnitude smaller than any energy scale possible in the system
or even the thermal broadening energy scale in the experiment. Initially
it was thought \cite{ye} that may be there is some accidental charge
addition into the system. For example it was argued \cite{ye} that the
states of the ring may play a role if the states of the dot are coupled to
the states of the ring. It was shown that charging of a ring state or
addition of a charge in the ring can produce a sudden phase change also
which can be misinterpreted as a phase change due to the dot. But repeated
experiments showed that this sharp phase change is a very general feature
occurring between every resonances and one can not take spurious charging
effects to be a logical explanation. So it was felt that one has to find a
new mechanism for this phase change that does not fit with the theoretical
frame work of Friedel sum rule and Bright-Wigner resonance. We used the
discontinuous phase change at the transmission zeroes of the stub
structure to give such a new mechanism \cite{deo4}. The stub was used as
theoretical model and the exact geometry of the stub is not important
\cite{deo5}. The essential requirement is the degeneracy between a
scattering state and a resonance state that yields a Fano resonance
\cite{deo5}. Since the quantum dot can support bound states and in a
finite width wire these bound states are always degenerate with scattering
states, we have to understand the phase changes in terms of Fano
resonances \cite{deo5}. Each of the Fano resonances have a zero-pole
pair. Two consecutive resonances are largely separated by the charging
energy of the dot. And hence there will be a phase change of $\pi$ at the
pole according to Friedel sum rule because a charge is captured by the dot
and a phase change of $\pi$ at the zero that will lie between two well
separated poles. And hence this mechanism yields the basic feature of
systematic phase changes as observed in the experiment. It also explains
the extremely small energy scale over which the phase change between the
poles occur. This explanation was further supported by Refs.
\cite{ryu2,lee,tan}. A similar case was studied in Ref. \cite{wu} where
they show the transmission zeroes and abrupt phase changes arise due to
degeneracy of ``dot states'' with states of the ``complementary part'' and
hence these are also Fano type resonances.

\section{Some consequences of parity violation}

In this section we will briefly discuss the results of Ref.  \cite{mos}.
Essentially it can be shown that break down of parity effect in the
ring-stub system results in very non-trivial temperature dependence of
persistent currents, a feature that may be exploited to experimentally
study the effects of temperature on a quantum mechanical phenomenon like
persistent currents.

We consider both the grand canonical case (When the particle exchange with
a reservoir at temperature $T$ is present and the reservoir fixes the
chemical potential $\mu$. In this case we will denote the persistent
current as $I_\mu$) and the canonical case (When number $N$ of particles
in the ring-stub system is conserved. In this case we will denote the
persistent current as $I_N$).  For the grand canonical case we suppose
that the coupling to a reservoir is weak enough and the eigenvalues of
electron wave number $k$ are not affected by the reservoir \cite{che}.
They are defined by the following equation \cite{deo2}

   \begin{equation}
    \cos(\alpha)=0.5\sin(ku)\cot(kv)+\cos(ku),
    \label{Eq1}
   \end{equation}
   \\
   \noindent

where $\alpha=2\pi\phi/\phi_0$, with $\phi_0=hc/e$ being the flux quantum.
Note, that Eq. (\ref{Eq1}) is obtained under the Griffith boundary
conditions \cite{gri} which take into account both the continuity of an
electron wave function and the conservation of current at the junction of
the ring and the stub; and hard wall boundary condition at the dead end of
the stub. Each of the roots $k_n$ of Eq.(\ref{Eq1}) determines the
one-electron eigenstate with an energy $\epsilon_n=\hbar^2k_n^2/(2m)$ as a
function of the magnetic flux $\phi$. Further we calculate the persistent
current $I_{N/\mu}=-\partial F_{N/\mu}/\partial \phi$ \cite{bye}, where
$F_N$ is the free energy for the regime $N=const$ and $F_\mu$ is the
thermodynamic potential for the regime $\mu=const$. In the latter case for
the system of noninteracting electrons the problem is greatly simplified
as we can use the Fermi distribution function
$f_0(\epsilon)=(1+\exp[(\epsilon-\mu)/T] )^{-1}$ when we fill up the
energy levels in the ring-stub system and we can write the persistent
current as follows \cite{che}

   \begin{equation}
    I_\mu=\sum_n I_n f_0(\epsilon_n),
    \label{Eq2}
   \end{equation}
   \\
   \noindent

where $I_n$ is a quantum mechanical current carried by the $n$th level and
is given by \cite{deo2}

   \begin{equation}
    \frac{\hbar I_n}{e}=\frac{2k_n\sin(\alpha)}{\frac{u}{2}\cos(k_nu)
   \cot(k_nv)-[\frac{v}{2}{\rm cosec}^2(k_nv)+u]\sin(k_nu)}.
    \label{Eq3}
   \end{equation}
   \\
 For the case of $N=const$ we must calculate the partition
   function $Z$ which determines the free energy  $F_N=-T\ln(Z)$

   \begin{equation}
    Z=\sum_m \exp\left( -\frac{E_m}{T} \right),
    \label{Eq4}
   \end{equation}
   \\
where $E_m$ is the energy of a many electron level. For the system of $N$
spinless noninteracting electrons $E_m$ is a sum over $N$ different
(pursuant to the Pauli principle) one-electron energies
$E_m=\sum_{i=1}^{N} \epsilon_{n_i}$, where the index $m$ numbers the
different series $\{\epsilon_{n_1},...,\epsilon_{n_N}\}_m$, where $n$ is
level index and $i$ is the particle index.  For instance, the ground state
energy is $E_0=\sum_{n=1}^{N}\epsilon_{nn}$.

We show a non-monotonous temperature dependence of the persistent currents
in this ballistic ring-stub system in the grand canonical ($I_\mu$) as
well as in the canonical case ($I_N$).  There is a crossover temperature
$T^*$, below which persistent currents increase in magnitude with
temperature while it decreases above this temperature. This is in contrast
to persistent currents in rings being monotonously affected by
temperature. $T^*$ is parameter dependent but of the order of
$\Delta_u/\pi^2k_B$, where $\Delta_u$ is the level spacing of the isolated
ring. For the grand canonical case $T^*$ is half of that for the canonical
case. We also show that such a non-monotonous temperature dependence can
naturally lead to a crossover from $\phi_0/2$ periodicity to $\phi_0$
periodicity of the persistent currents as a function of temperature, where
$\phi_0$=hc/e. This is essentially because each of the harmonics can show
an enhancement with temperature, at low temperatures, the crossover
temperature for the $m$th harmonic being approximately $T^*/m$. Hence the
second harmonic can peak at a lower temperature than the first and so can
overtake the first harmonic in a certain temperature window. The
temperature dependence of the fundamental periodicity is a very unique
feature that cannot be seen in the ring system. The ring system can under
certain circumstances show the $\phi_0$ periodicity as well as the
$\phi_0/2$ periodicity, but the periodicity remain unchanged with
temperature. In the ring-stub system temperature
enhancement of the first harmonic can be very robust and experimentally
one can observe each of the harmonics separately.

To summarize, the ring-stub system has a lot of non-trivial temperature
dependence of persistent currents which can provide some experimental
motivation. The new phase discussed in the last section is the key source
of the non-trivial temperature dependence.

\section{Dephasing of AB-oscillations due to spin-flip scattering in
  one arm of the AB-ring}

Understanding of dephasing and decoherence in mesoscopic systems connected
with measurement and the quantum to classical transition is a subject of
current interest. The importance of decoherence in realizing the classical
world is well known \cite{zur}. The loss of interference (dephasing) in
mesoscopic rings from the point of view of the trace left by the
interfering particle on its environment or the effect of environment on
the phase of the electron wavefunction has been clearly discussed in a
seminal paper by Stern et. al. \cite{sai}.  Our work \cite{sah} is in the
close spirit to the above mentioned work of Stern et. al.. We have studied
the transmission across a single channel AB-ring geometry using the
quantum waveguide theory. In one of the arms (say upper arm) we have
incorporated a spin half (S=1/2) magnetic impurity. Electron interacts
with this impurity via an exchange interaction of the type $J~S.s$, where
$s$ is the spin of the electron and $J$ is the coupling strength. This
interaction conserves the total spin ${\cal S}=S+s$. The magnetic impurity
does not have its own dynamics. Depending on the initial direction of
incident electron spin and that of the spin of the impurity spin-flip
scattering takes place without exchange of any energy. Consider two
partial waves going across once and initial spin is up, i.e., electron is
in $s_z=+1/2$ state with magnetic impurity being in down ($S_z=-1/2$)
state. In this case the electron can get spin-flipped while traversing the
upper arm, but no spin-flip occurs while traversing the lower arms. Thus,
we would naively expect transmission amplitude for spin down case to
vanish due to destructive interference. This is because looking at the
spin of the magnetic impurity we would know which path the electron has
taken. This is in close spirit to which-path detector or information
models being studied in the quantum theory of measurement. However, this
naive expectation turns out to be incorrect. This is because even after
getting spin-flipped in the upper arm the electron may get reflected and
finally traverse the lower arm and contribute to the spin down component
of transmission. This study reveals several interesting features such as
spin polarized transport (spin conductance or polarization) is asymmetric
in magnetic field whereas two probe charge conductance is symmetric. There
is no systematics in the harmonic components in flux of spin down and spin
up components of transmission coefficient (owing to the sensitivity to the
details of geometry and scattering strength). However, we have noticed
that the amplitude of oscillations of the transmission coefficient
(visibility)  for the down spin is small as compared to the up spin. This
clearly brings out the feature of dephasing in this simple model.

\section{Conclusion}

In conclusion, we have discussed above our own work on several of
phenomenon arising basically due to quantum interference effects in the
mesoscopic systems. For this purpose we have considered single channel
case and various geometries. We have worked in the framework of free
electron model. We have also suggested the means of probing the
experimental manifestations of these phenomenon and their applications. A
careful study of these phenomena in presence of inelastic scattering due
to phonons and electron-electron interaction is called for.

{\bf Figure captions}

\noindent {\bf Fig.~1 (a)} The thin lines represent single channel
quantum wires supporting a ring. Through the center
of the ring there is a magnetic flux $\phi$ perpendicular to the
plane of the paper. The reservoir-1 is at a chemical potential
$\mu_1$ and the reservoir-2 is at a chemical potential $\mu_2$.
There is a delta function repulsive potential of strength $V$
at the site marked X.

\noindent {\bf Fig.~1 (b)} The thin lines represent single channel
quantum wires along with a ring. Through the center
of the ring there is a magnetic flux $\phi$ perpendicular to the
plane of the paper. The reservoir-1 is at a chemical potential
$\mu_1$ and the reservoir-2 is at a chemical potential $\mu_2$.
There is a delta function repulsive potential of strength $V$
at the site marked X. The two archs of the ring are of length
$L_1$ (upper arch) and $L_2$ (lower arch). Current through
$L_1$ is $I_1$ and that through $L_2$ is $I_2$.

\noindent {\bf Fig.~2} An ideal ring of length $L$
connected to an electron
reservoir by an ideal wire. Along the thin lines the quantum
mechanical potential is zero and along the thick lines it is $V$.

\noindent {\bf Fig.~3} A quantum ring connected freely to a
finite quantum wire. Length of the the ring is $u$ and length
of the wire is $v$.

\noindent {\bf Fig.~4 (a)} Graphical solutions for the allowed
modes in a ring-stub geometry for $v/u$=0.2

\noindent {\bf Fig.~4 (b)} Graphical solutions for the allowed
modes in a ring-stub geometry for $v/u$=0.21
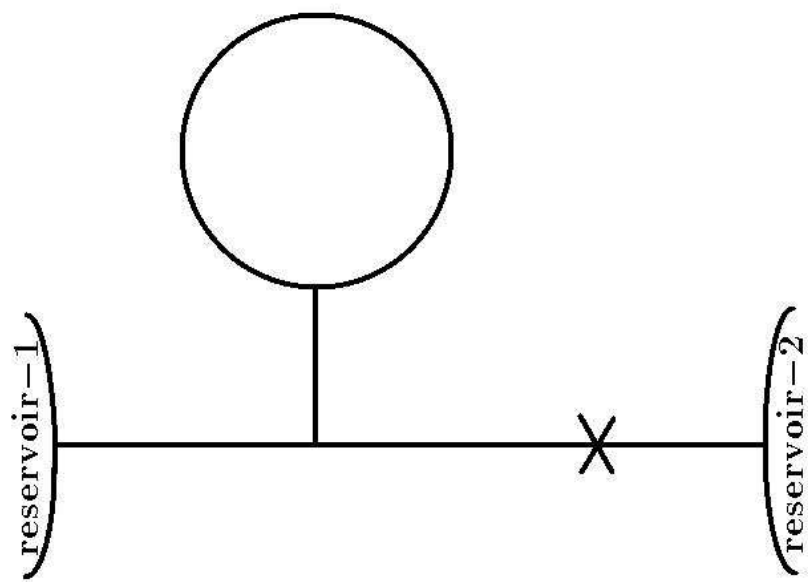
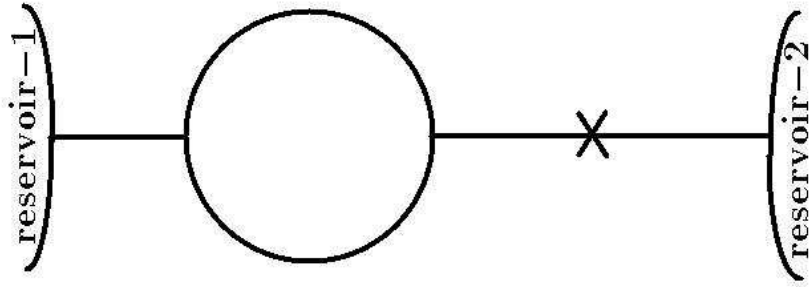
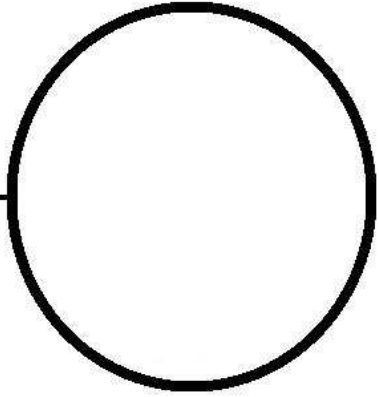
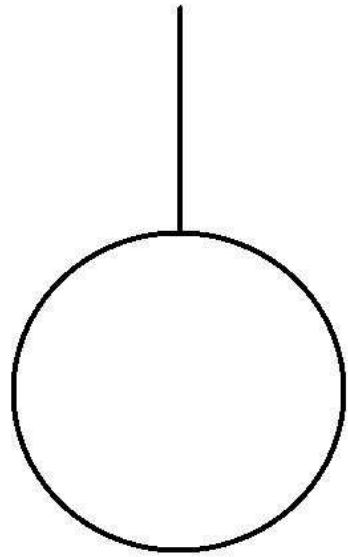
\end{document}